\documentclass{bmcart}

\usepackage[utf8]{inputenc} 
\usepackage{url}
\usepackage{graphicx}
\usepackage{booktabs}
\usepackage{multirow}

\startlocaldefs
\endlocaldefs

\begin{document}

\begin{frontmatter}

\begin{fmbox}
\dochead{Research}

\title{Quantifying the Economic Impact of Extreme Shocks on Businesses using Human Mobility Data: a Bayesian Causal Inference Approach}

\author[
  addressref={aff2},              
  email={tyabe@purdue.edu}   
]{\inits{TY}\fnm{Takahiro} \snm{Yabe}}
\author[
  addressref={aff2},                   
  email={zhan2854@purdue.edu}   
]{\inits{YZ}\fnm{Yunchang} \snm{Zhang}}
\author[
   addressref={aff2},
   corref={aff2},  
   email={sukkusur@purdue.edu}
]{\inits{SVU}\fnm{Satish V} \snm{Ukkusuri}}

\address[id=aff2]{
  \orgname{Lyles School of Civil Engineering, Purdue University}, 
  \street{550 Stadium Mall Avenue},                     %
  \city{West Lafayette},                              
  \cny{USA}                                    
}


\end{fmbox}

\begin{abstractbox}

\begin{abstract} 
In recent years, extreme shocks, such as natural disasters, are increasing in both frequency and intensity, causing significant economic loss to many cities around the world.
Quantifying the economic cost of local businesses after extreme shocks is important for post-disaster assessment and pre-disaster planning. 
Conventionally, surveys have been the primary source of data used to quantify damages inflicted on businesses by disasters. 
However, surveys often suffer from high cost and long time for implementation, spatio-temporal sparsity in observations, and limitations in scalability. 
Recently, large scale human mobility data (e.g. mobile phone GPS) have been used to observe and analyze human mobility patterns in an unprecedented spatio-temporal granularity and scale. 
In this work, we use location data collected from mobile phones to estimate and analyze the causal impact of hurricanes on business performance. 
To quantify the causal impact of the disaster, we use a Bayesian structural time series model to predict the counterfactual performances of affected businesses (\textit{what if the disaster did not occur?}), which may use performances of other businesses outside the disaster areas as covariates. 
The method is tested to quantify the resilience of 635 businesses across 9 categories in Puerto Rico after Hurricane Maria.
Furthermore, hierarchical Bayesian models are used to reveal the effect of business characteristics such as location and category on the long-term resilience of businesses. 
The study presents a novel and more efficient method to quantify business resilience, which could assist policy makers in disaster preparation and relief processes. 
\end{abstract}

\begin{keyword}
\kwd{disaster resilience}
\kwd{mobile phones}
\kwd{human mobility}
\kwd{causal inference}
\end{keyword}


\end{abstractbox}

\end{frontmatter}

\section*{Introduction}
Recently, natural hazards are increasing both in frequency and intensity in many parts of the world. 
The economic losses caused by such extreme events exceeded a total of \$2.5 trillion across the globe since 2000, and are rising each year due to rapid urbanization in many cities \cite{UN}. 
With the intensifying threat of significant economic damage, improving the resilience of cities has attracted interest from a wide range of fields including public policy, urban planning, complex systems, and economics \cite{eakin2017opinion}.
Among the various dimensions of disaster resilience, the ability of businesses to bounce back is a critical component that significantly contributes to the economic recovery of cities after disasters. 

Previous studies have analyzed the post-disaster recovery of businesses through the means of surveys and interviews. 
Such studies have identified factors such as pre-disaster size of the business and category of business that partly explain the reopening and demise of businesses after disasters including Hurricanes Katrina \cite{marshall2015predicting,sydnor2017analysis}, Andrew \cite{webb2002predicting}, and more recently, Harvey \cite{lee2019business}. 
Although these studies provide a general understanding of the effect of various characteristics of businesses that affect the post-disaster recovery performances, they suffer from two critical drawbacks. 
First, observations are limited to discrete measurements at a few number of timings, failing to give a quantifiable, continuous and longitudinal understanding of the recovery process of businesses. 
Second, the applied methods fail to model the causal effect of the disaster, which require a statistical framework that predicts the performances of businesses if the disaster did not occur.

With the emergence of novel and often large-scale data collected from mobile sensors and online social platforms, we are now capable of observing and analyzing the dynamics of people, goods, and information at an unprecedented spatio-temporal granularity \cite{batty2013big}.
In particular, location data collected from mobile phones (e.g. call detail records, GPS trajectories) have enabled us to observe individual mobility patterns at an unprecedented high spatio-temporal granularity \cite{gonzalez2008understanding,blondel2015survey}. 
Such datasets are now utilized for a wide range of applications to solve urban challenges including population density estimation \cite{deville2014dynamic, wardrop2018spatially}, traffic estimation \cite{iqbal2014development,calabrese2011estimating}, predicting poverty \cite{blumenstock2015predicting}, and modeling spread of epidemics \cite{bengtsson2015using}. 
In the context of extreme events, several studies have used mobile phone data to analyze the mobility patterns during and after disasters such as earthquakes \cite{lu2012predictability,song2014prediction,wilson2016rapid}, cyclones \cite{lu2016unveiling}, and other anomalous events \cite{bagrow2011collective}. 
Despite such progress, none of the previous studies have used large scale mobility data to analyze the recovery of businesses after disasters. 

Recent advances in statistical models, in particular Bayesian structural time series (BSTS) models, allow flexible predictions of time series data, which can be used to estimate the causal impact \cite{harvey2014structural}.  
BSTS has several advantages over conventional difference in differences models \cite{bellman1963differential}, including its flexibility to model the causal impact over a longitudinal time horizon rather that across 2 time points.
A recent study using website click-through data applied BSTS models to quantify the causal impact of an online advertisement \cite{brodersen2015inferring}. 
We take advantage of this recently proposed methodology to quantify the causal impact of hurricanes on businesses in Puerto Rico. 

\begin{figure}
    \centering
    \includegraphics[width=\textwidth]{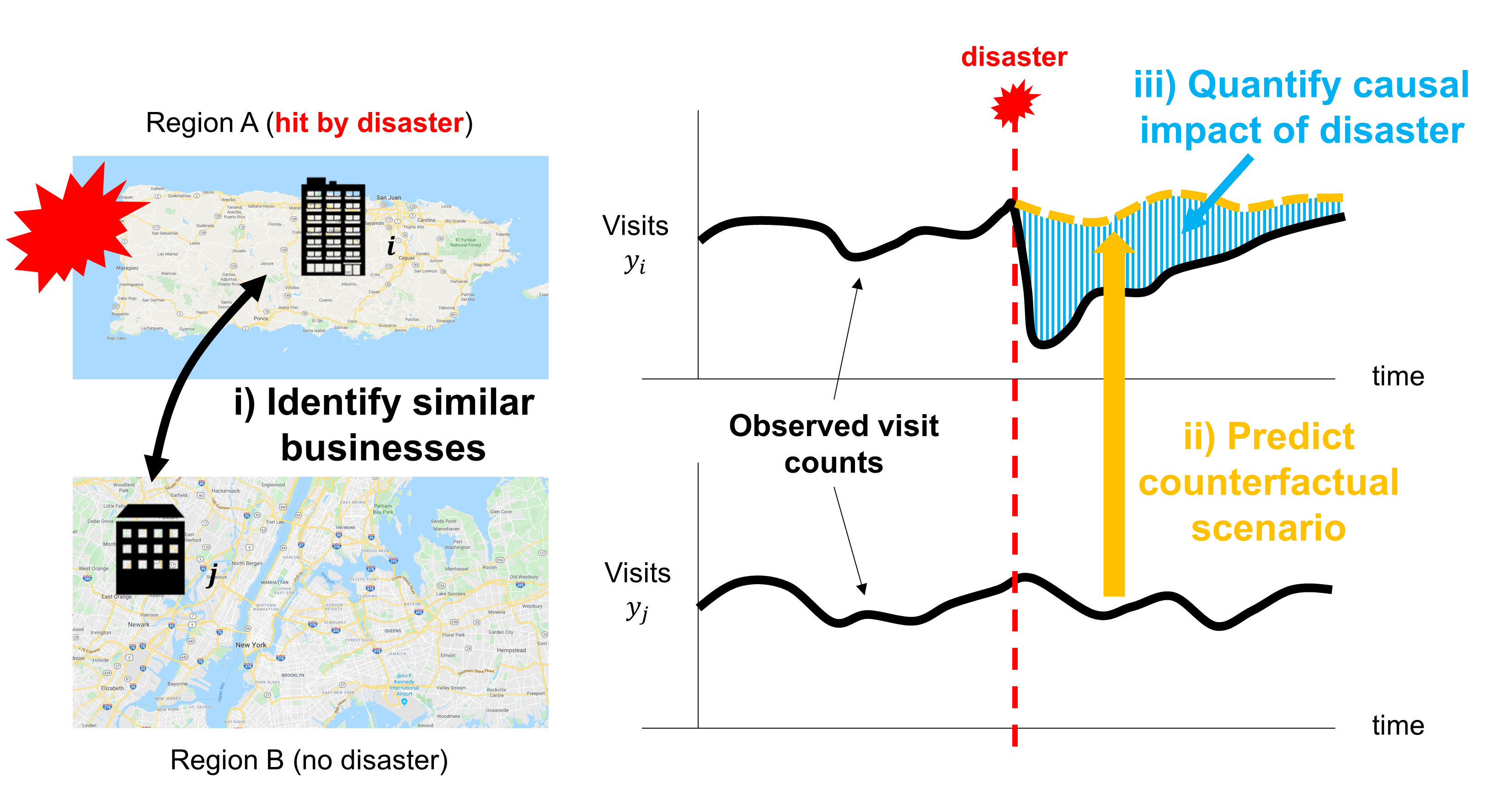}
    \caption{\textbf{Overview of study.} Our causal inference procedure is composed of 3 steps. i) To measure the causal impact of the disaster on business $i$, we first identify a similar business $j$ in another region which was not affected by the disaster. ii) We then predict the counterfactual (``what-if the disaster did not occur?'') visit count of $i$ after the disaster timing using observed data from $j$. iii) We can quantify the causal impact of the disaster by taking the difference between the predicted and observed visit counts in $i$.}
    \label{overview}
\end{figure}

This study makes several contributions to overcome the aforementioned drawbacks in the previous studies on business recovery after disasters. 
First, this is the first work to utilize large scale mobility data collected from mobile phones to estimate the popularity of businesses before, during and after a disaster. 
Second, a Bayesian structural time series model combined with an inter-city matching scheme is proposed to infer the causal impact of the disaster on businesses. 
Third, the proposed methodology is applied on mobile phone data collected from Puerto Rico to quantify the resilience of businesses after Hurricane Maria. 
Figure \ref{overview} illustrates the overview of the study. 
The causal inference procedure is composed of 3 steps. i) To measure the causal impact of the disaster on business $i$, we first identify a similar business $j$ in another region which was not affected by the disaster. ii) We then predict the counterfactual (\textit{``what-if the disaster did not occur?''}) visit count of $i$ after the disaster timing using observed data from $j$, via a Bayesian structural time series model. iii) As a result, we can quantify the causal impact of the disaster by taking the difference between the predicted and observed visit counts in $i$.

\section*{Related Works}
\subsection*{Resilience of businesses after disasters}
The economic impact of disasters on businesses have conventionally been studied through surveys that are performed after the disaster. 
Studies using surveys have identified various factors that affect the reopening and demise of businesses after disasters through econometric models (e.g. logistic regression) \cite{marshall2015predicting,sydnor2017analysis}.
Important factors that affected the outcomes of businesses after Hurricane Katrina include household size of the business owner, previous disaster experience, number of employees, business age, and legal structure of the business \cite{marshall2015predicting}. 
The qualitative details in the collected data are a significant advantage of surveys. 
However, surveys suffer from various drawbacks such as the high cost and long time for implementation, spatio-temporal sparsity in observations, and limitations in scalability.
Due to these limitations, it is difficult to obtain a quantifiable, continuous and longitudinal understanding of the recovery process of businesses. 
Moreover, the applied methods fail to model the causal effect of the disaster, which require a statistical framework that predicts the performances of businesses if the disaster did not occur.

\subsection*{Mobility analysis using mobile phone data}
With the emergence of novel and often large-scale data collected from mobile sensors and online social platforms, we are now capable of observing and analyzing the dynamics of people, goods, and information at an unprecedented spatio-temporal granularity \cite{batty2013big}.
In particular, location data collected from mobile phones (e.g. call detail records, GPS trajectories) have enabled us to observe individual mobility patterns at an unprecedented high spatio-temporal granularity \cite{gonzalez2008understanding,blondel2015survey}. 
These new datasets are becoming new standards for population level studies, and are used to understand the population distribution in cities \cite{deville2014dynamic}. 
Such datasets are now utilized for a wide range of applications to solve urban challenges including population density estimation \cite{wardrop2018spatially}, estimation of dynamic traffic flows \cite{iqbal2014development,calabrese2011estimating}, predicting poverty in developing counties \cite{blumenstock2015predicting}, and modeling the impact of human mobility patterns on the spread of epidemics \cite{bengtsson2015using}. 
In the context of extreme events and disasters, several studies have used mobile phone data to analyze the mobility patterns during and after disasters \cite{lu2012predictability,song2014prediction,wilson2016rapid}. 
Studies using such large scale data has revealed important insights on the evacuation and migration patterns of the affected people \cite{lu2012predictability, lu2016unveiling}. 
Despite such progress, none of the previous studies have used large scale mobility data to analyze the recovery of businesses after disasters. 
A recent study using mobile phone GPS data (same data used in this study) revealed the impact of the recent policy regarding the usage of bathrooms in Starbucks on the visit behavior of people to the cafe chain \cite{gurun}. 
They validated that the spatio-temporal granularity of the mobile phone GPS data is of sufficient detail to analyze the store level visit behavior. 
In this study we apply a similar approach, and estimate the visit behavior of people to stores and businesses using mobile phone GPS data.

\subsection*{Statistical methods for causal inference}
\subsubsection*{Difference in Differences}
Difference in differences (DiD) method is a statistical method estimating the treatment effects between the "treatment" group versus the "control" group. For a specific before-and-after study, DiD compares the average change over time between treatment and control groups, which provides us a classical method in estimating causal effects of natural experiments without strictly randomization \cite{bellman1963differential, goodman2018difference}. However, classical DiD have several limitations: 1. It follows the parallel trends assumption that requires the differences between treatment and control group are invariant overtime in absence of the treatment \cite{abadie2005semiparametric, antonakis2010making}. In a before-and-after study, the parallel trends assumption necessitates the dynamics of the means for the two groups should be balanced overtime. Consequently, issues such as time-correlated responses will contaminate the causal inference with DiD \cite{hansen2007asymptotic}. 2. Only two time steps - pre-treatment time and post-treatment time are considered in the classical DiD that merely captures the static causal effects for a specific before-and-after study. It can be implausible and useless if the outcome of interest dynamically changes over time such as recovery patterns after disaster, radioactive decay etc \cite{brodersen2015inferring}.

\subsubsection*{Bayesian Structural Time Series Models}
Compared with the classical DiD model, a structural time series model promisingly relaxes the parallel trends assumption and captures the variations of time-varying local trends and seasonality for time-correlated response variables \cite{harvey2014structural, harvey1990forecasting}. In addition, structural time series models encompass a flexible model structure that enables us to analyze the dynamic effects of the outcome of interest during a time period \cite{leeflang2009creating}.  
Due to a large number of predictors in structural time series models, a Bayesian approach was introduced to sparse the estimation of coefficients. 
Scott and Varian \cite{scott2013predicting, scott2013bayesian} proposed a spike-and-slab prior to the regression coefficients in a Google search query study, which significantly reduces the size of the problem. 
Nakajima and West  \cite{nakajima2013bayesian} elicited a dynamic spike-and-slab prior that sparsified the estimation of time-varying parameters for a Bayesian macroeconomic time series model. 
The most recent Google study for causal inference of a market intervention \cite{brodersen2015inferring} slightly revised the dynamic version of pike-and-slab prior \cite{nakajima2013bayesian} with a weakly informative prior. 
In addition, the Bayesian structural time series models (BSTS) have been constructed to strengthen causal inference for time series data. To address the fundamental problem in causal inference \cite{imbens2017rubin}, pre-treatment observations are trained and tested via BSTS and consequently the fitted BSTS can simulate the counterfactual as the synthetic post-treatment controls via posterior predictive samples. 
This method is extensively applied in causal inference throughout various fields, such as socio-economics \cite{harvey2007trends, poyser2019exploring}, political science \cite{brandt2006advances, brandt2011real}, environmental studies \cite{jiang2013very, de2016inferring}.

The causal inference methodology proposed in the previous section is applied on data collected from Puerto Rico before and after Hurricane Maria, which made landfall on September 20th, 2019, and caused a long term devastating humanitarian and economic crisis. 
Fatalities as a consequence of Maria are still under investigation, however recent estimates suggest that between 793 to 8,498 excess deaths occurred following the storm \cite{kishore2018mortality}. 
Heavy rainfall, flooding, storm surge, and high winds caused considerable damage to various infrastructure systems, causing power outages and water shortages for the entire island for months. 
Total economic losses to Puerto Rico and the US Virgin Islands are estimated to be \$90 billion, with a 90\% confidence range of $\pm$\$25.0 billion, which makes Maria the third costliest hurricane in U.S. history, behind Katrina (2005) and Harvey (2017) \cite{pasch2018national}.

Three main data sources are used in this study: (1) business visit data collected from mobile phones, (2) spatial distribution of housing damages due to Hurricane Maria, and (3) socio-economic factors of census blocks in Puerto Rico. 
In this section, we describe how these datasets were collected, processed, and used to infer the causal impact of the hurricane on businesses. 

\subsection*{Business visit data collected from mobile phones}
Establishment-level visit data are provided by Safegraph\footnote{\url{https://www.safegraph.com/}}, a company that aggregates anonymized location data collected from smartphone applications to provide insights about physical places. 
Safegraph's location dataset covers around 10\% of all smartphones in the United States, and each observation is consisted of a unique (but anonymized) user ID, longitude, latitude, and timestamp information. 
The longitude and latitude information are accurate to within a few meters, allowing us to analyze the visit counts to each establishment. 
To detect a user visiting an establishment, the location data are first cleaned by removing GPS signal drifts and jumpy observations using a spatial threshold, then clustered into a staypoint using a spatio-temporal DBSCAN algorithm. 
Then, the visited establishment is predicted from establishments nearby the clustered staypoint by using a machine learning algorithm that takes into account various features such as distances from establishment to the cluster centroid, time of day, and North American Industry Classification System (NAICS) code. 
Performing this procedure for all days in the dataset produces a time series data of daily visit counts for each establishment. 

We use daily visit data of establishments located in Puerto Rico and the State of New York between January 2017 and March 2018 to quantify the causal impact of the hurricane on business resilience. 
Daily visit data of businesses in New York are used since these businesses constitute a reasonable control group which were not affected by the disruptions caused by Hurricane Maria. 
How we use the visit data from the control group in the causal inference model is explained in the Methods section. 
We limit the analysis to business categories that sell products or services directly to the customers, since we will approximate business performances from the number of visits per day, observed from mobile phone data. 
We also limit the analysis to medium or large sized businesses with more than 100 customers per day on average (before the disaster), since we are not able to observe visit patterns below that level using mobile phone data. 
As summarized in Table \ref{table1}, daily visit data of a total of 635 businesses in Puerto Rico were analyzed, along with 1,102 and 10,409 businesses in Manhattan and Up-State New York, respectively.

\begin{table}
\caption{Summary statistics of business visit time series data.}
      \begin{tabular}{lccc}
        \toprule
        \multirow{2}{*}{Business Category} & \multicolumn{3}{c}{Region} \\
        \cmidrule{2-4}
        & Puerto Rico & Downstate New York & Upstate New York  \\ \midrule
        Building Material & 62 & 36 & 584 \\
        Gasoline Stations & 10 & 143 & 1,160 \\ 
        Grocery Stores & 34 & 97 & 1,227   \\
        Hospitals & 12  & 25 &  76 \\
        Hotels & 8 & 81 & 692   \\
        Restaurants & 322 & 585 & 6,352  \\
        Supermarkets & 61 & 0 & 52  \\
        Telecommunication & 101 & 9 & 67   \\
        Universities & 25  & 34 & 199  \\
        \midrule
        Total & 635 & 1,102 & 10,409 \\
        \bottomrule
      \end{tabular}
      \label{table1}
\end{table}

\begin{figure}
    \centering
    \includegraphics[width=0.9\textwidth]{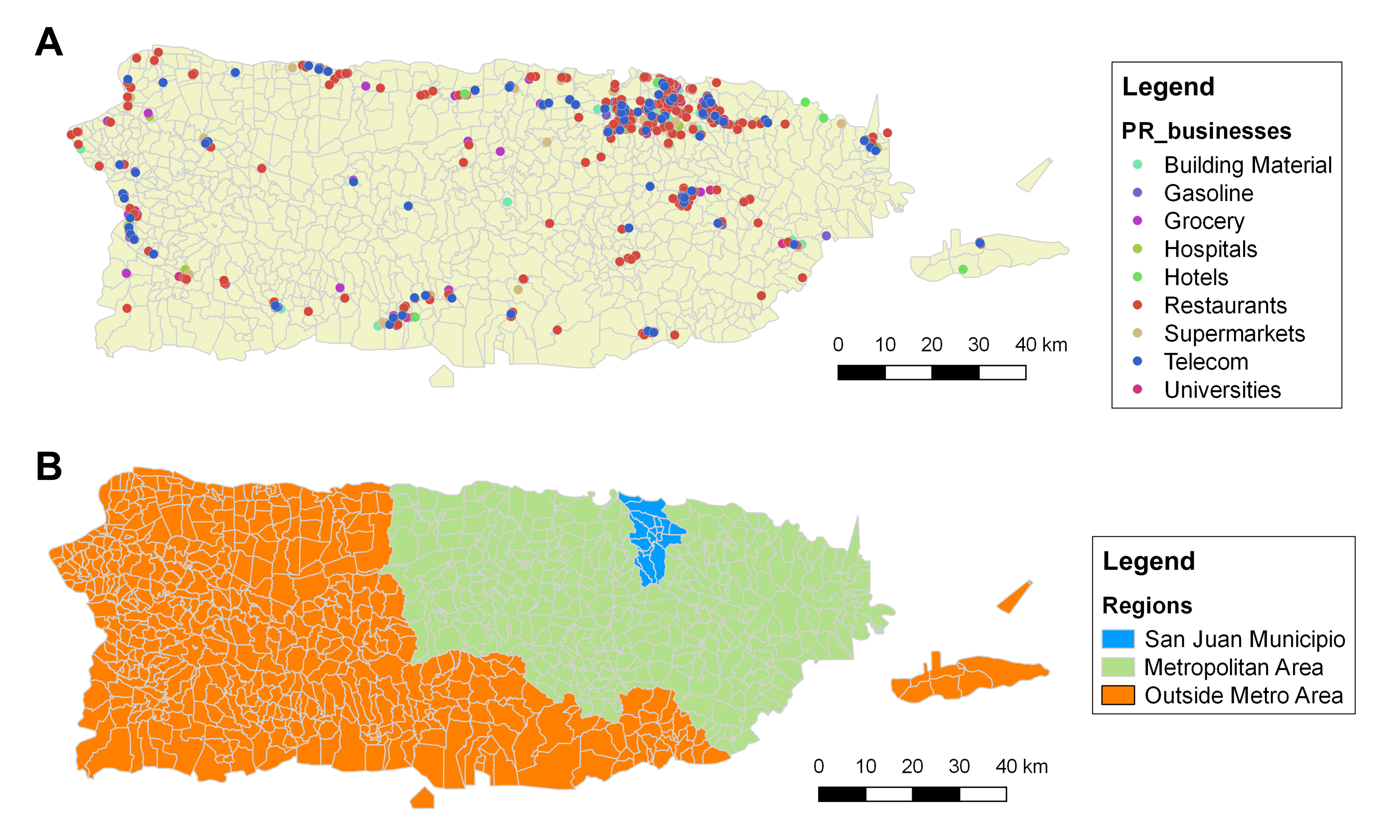} 
    \caption{\textbf{Characteristics of businesses in Puerto Rico.} (A) Business locations and categories in Puerto Rico. (B) 3 regions of Puerto Rico used in this study.}
    \label{map}
\end{figure}

\subsection*{Socio-economic data}
In this study, population and income data of each county were used for later analysis. 
Population data were obtained from the US National Census\footnote{\url{https://www.census.gov/}}, and median income data were obtained from the American Community Survey\footnote{\url{https://www.census.gov/programs-surveys/acs}}. 

\subsection*{Spatial distribution of housing damages due to Hurricane Maria}
Physical damage caused by the hurricane are measured by the housing damage rates in each county, which was provided through the ``Housing Assistance Data'' provided by the Federal Emergency Management Agency (FEMA). 
The raw data can be found through the link\footnote{\url{https://www.fema.gov/media-library/assets/documents/34758}}.
We defined ``housing damage rate'' for each county as the total number of houses that were inspected to have had more than \$ 10,000 worth of damage due to the target hurricane, divided by the number of households in that county. 
Many of the counties in Puerto Rico experienced high housing damage rates, between 20\% and 60\%.

\section*{Methods}
\subsection*{Bayesian structural time series model}
The basic structural time series model is defined as the following:
\begin{eqnarray}
    \left\{
    \begin{array}{c}
    y_{t,i} = \mu_{t,i} + \tau_{t,i} + \beta x_{t,i} + \epsilon_{t,i} \ \ \ \ \forall t\\
    \epsilon_{t,i} \sim \mathcal{N}(0, \sigma_y^2) \\
    \sigma_y \sim Cauchy(0, 2.5)
    \end{array}
    \right.
\end{eqnarray}
where $y_{t,i}$ is the observed daily visits to business $i$ on day $t$ in the target region (in our case, Puerto Rico). 
$y_{t,i}t$ is predicted by state components $\mu_{t,i}$, $\tau_{t,i}$ and $\beta x_{t,i}$ that capture critical features of the time-series data \cite{brodersen2015inferring}. 
A weakly informative prior is elicited for each state component. 

\textbf{Local Level Trend: }The local level model represents local variations of the time series data. 
To simplify the model structure, we assume the mean of the trend is a random walk with the initialization of $\mu_1$:
\begin{eqnarray}
    \left\{
    \begin{array}{c}
    \mu_{t+1,i} = \mu_{t,i} + \eta{_{1,t,i}}\ \ \ \ \  \forall t>1\\
    \mu_{1,i}\sim \mathcal{N}(\mu_0, \sigma_0^2) \\
    \eta_{1,t,i} \sim \mathcal{N}(0, \sigma_\mu^2) \\
    \sigma_0, \mu_0, \sigma_\mu \sim Cauchy(0,2.5)
    \end{array}
    \right.
\end{eqnarray}

\textbf{Seasonality: } Let $S$ denote the total number of seasons.
The sum of seasonal effects over $S$ time periods is assumed to be zero. 
In this study, weekly seasonality is taken into account ($S = 7$) with the initialization of $\tau_{1,i}$, $\tau_{2,i}$, $\tau_{3,i}$, $\tau_{4,i}$, $\tau_{5,i}$, and $\tau_{6,i}$:
\begin{eqnarray}
    \left\{
    \begin{array}{c}
    \tau_{t+1,i} = - \sum_{s=1}^S \tau_{t-s,i} + \eta_{2,t} \ \ \ \forall t>1 \\
    \tau_{1,i},\tau_{2,i}, \tau_{3,i}, \tau_{4,i}, \tau_{5,i}, \tau_{6,i} \sim \mathcal{N}(\mu_{\tau_0}, \sigma_{\tau_0}^2) \\
    \eta_{2,t} \sim \mathcal{N}(0, \sigma_\tau^2) \\
    \mu_{\tau_0}, \sigma_{\tau_0}, \sigma_\tau \sim Cauchy(0, 2.5)
    \end{array}
    \right.
\end{eqnarray}

\subsubsection*{Choice of Covariates}
Apart from the local level model and seasonality, there are other unobserved effects such as impacts of holidays and sport events that may contaminate the estimation of the $y_{t,i}$. 
To capture the unobserved heterogeneity, $x_{t,i}$ in Equation (1) is used as the simultaneous daily visits to a similar business type at time $t$ in a different region that was not affected by the disaster (in our case, New York). 
$x_{t,i}$ accounts for the shared variance of the time series data from two different regions. 
The static coefficient $\beta$ represents the relationship between daily visits to a specific business type from Puerto Rico and New York.
In this study, we test three methods for the choice of covariates, which we will test in the experiments Section: (i) no covariate, (ii) use the average daily visit trends of the same brand businesses in the other city as covariate (e.g. if $y_i$ was a Starbucks, we would use the average daily visit counts of all Starbucks in New York as the covariate), which we denote as $x_{category}$, and (iii) use the daily visit count of a specific business which has the highest correlation with the target business, which we denote as $x_{specific}$. For (iii), we compute the Pearson's correlation between the daily visit count data of the target business with that of all same category businesses in New York, and use the business with the highest Pearson R. 

\begin{figure}
\centering
    \includegraphics[width=\textwidth]{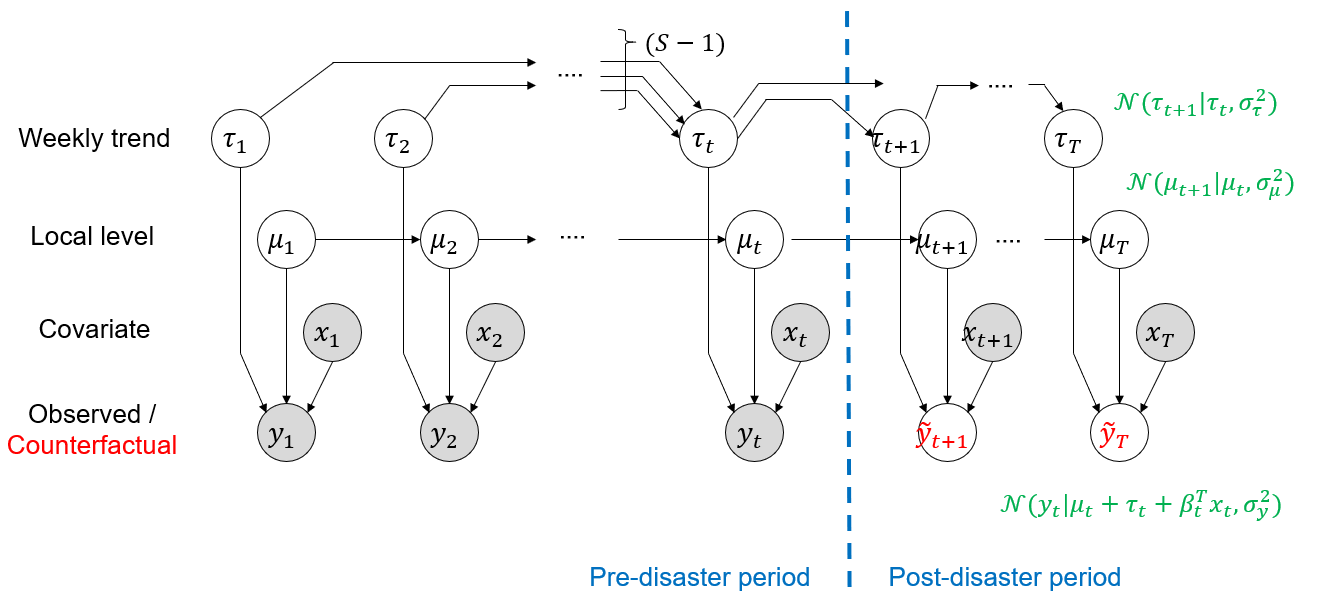}
    \caption{Graphical representation of the Bayesian structural time series model.}
\end{figure}

\subsection*{Estimating causal impact of disasters on businesses}

Let $N$ denote the total number of days observed. We first fit the BSTS model with pre-disaster data ($n$ = 150) from New York and Puerto Rico. 
For each business with index $i$, posterior predictive samples can be simulated to develop a counterfactual as the synthetic control group ($t = n+1,...,N$) from Equation (4).

\begin{equation}
    \hat{y_{t,i}} \sim p(\hat{y_{t,i}}|y_{t,i}) \ \ \ \ t \geq n
\end{equation}

Let $m \in [n, N]$ denote the the day when the Hurricane Maria struck Puerto Rico. 
Point-wise comparisons estimate the impacts of hurricane on daily visits to a target business type between treatment and control groups.

\begin{equation}
    \phi_{t,i} = \frac{y_{t,i} - \hat{y_{t,i}}}{\bar{y_i}} \ \ \ \ \ t = m+1,...,N
\end{equation}
where, $\bar{y_i}$ denotes the mean visit count to the visits prior to the disaster $(t < n)$. 
The impact $\phi_{t,i}$ is a normalized measure of the disaster impact to the business. 
$\phi_{t,i}$ measures the number of business-as-usual days worth of impact (damage) the disaster inflicted on the business. 

Moreover, We hope to estimate the cumulative causal effects of hurricane on a target business type over time, which represents the resilience of business after hurricane. 
The cumulative sum of causal increments is a practical quantity when the response variable $y_{t,i}$ is measured over time.
We calculate the total impact of the disaster to business $i$ by the following equation.

\begin{equation}
    \phi_i = \sum_{t = m}^N \phi_{t,i}  
\end{equation}

The cumulative sum of causal increments can be further transformed into the estimated total economic loss by multiplying average spending in dollar(s) per customer.

\begin{figure}
    \includegraphics[width=\textwidth]{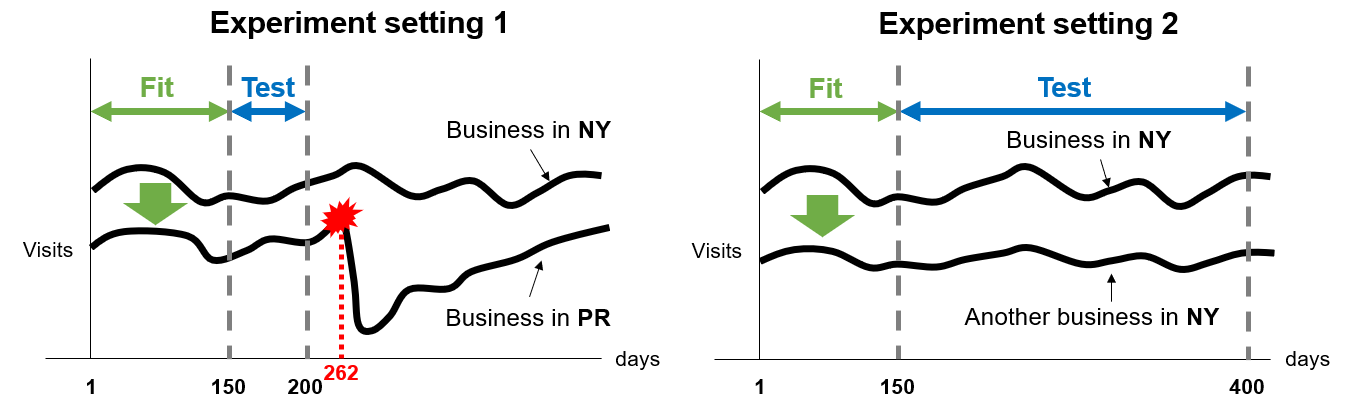}
    \caption{Experiment settings for model validation.}
    \label{setting}
\end{figure}

\section*{Model Validation}
\subsection*{Experiment Setup}
Daily visits to businesses in Puerto Rico and New York from January 2017 to March 2018 (400 days) are analyzed. 
As explained in the Methods Section, we will test three methods of selecting the covariate: no covariate, $x_{category}$, and $x_{specific}$. 
To verify the which type of covariate improves the prediction accuracy the most, two different model settings shown in Figure \ref{setting} will be explored: 
\begin{itemize}
    \item Setting 1 (Inter-State prediction): Pre-disaster data will be used from Puerto Rico and New York. 
    The model will be fitted using data until day 150, and tested using data between days 151 and 200.
    \item  Setting 2 (Intra-State prediction): To test the accuracy of long-term predictions, data from businesses in Manhattan will be used to predict the visit counts of businesses in Up-State New York, using the whole observation period (train: 0-150, test: 151-400).
\end{itemize}

\subsection*{Evaluation Metrics}
The prediction tasks will be evaluated using 2 different metrics: i) Pearson's R, which captures the correlation between the predicted and true time series values, and ii) mean absolute percentage error (MAPE), which captures the relative magnitude of the absolute error between the predicted and true time series values. 
MAPE is calculated by the following equation:
\begin{equation}
    MAPE_i = \frac{1}{n} \sum_{t=1}^n \left| \frac{y_{t,i} - \hat{y_{t,i}}}{y_{t,i}} \right| 
\end{equation}

We measure the performance of the methods using these two distinct metrics, where Pearson's R measures the relative correlation between the two sequences, while MAPE measures the absolute magnitude of discrepancy between the two vectors. 

\subsection*{Validation Results}
The performances of the BSTS models with three types of covariates, were tested on the aforementioned two experimental settings, using Pearson's correlation and MAPE as evaluation metrics. 
Table \ref{results} shows the performances of the three BSTS models on both settings. 
Surprisingly, although the model with business-category covariates perform the best on average in both experimental settings, the predictive performances of the three methods are quite similar.
Using extra covariates do not always improve the prediction model, and we see that over 34\% of the businesses in experiment setting 1 had best performances when not using extra covariates (similarly, over 40\% of businesses in experimental setting 2). 
Extra covariates, which are aimed to capture the long term trends and anomalies (e.g. New Years, Christmas), are not effective when making predictions of businesses that have less long-term variation and a relatively stable periodicity in visit counts. 
From experiment 1, we determine the best performing model out of the three for each business, and we use that business to predict the counterfactual daily visit counts after the disaster period. 

\begin{table}
\caption{Model validation results of two experimental settings.}
\begin{tabular}{ccccccc}
\toprule
 & & \multirow{2}{*}{\begin{tabular}{c}Evaluation\\Metric\end{tabular}}& \multicolumn{3}{c}{Use of Covariates} \\ 
\cmidrule{4-6}
 & & & No Covariates & $x_{category}$ & $x_{specific}$ \\ \midrule
\multirow{6}{*}{Setting 1} & \multirow{2}{*}{Train}  & MAPE & 12.35 ($\pm$16.67) & 10.50 ($\pm$14.03) & 10.66 ($\pm$14.46) \\ 
 & & Pearson R & 0.539 ($\pm$0.222) & 0.696 ($\pm$0.169) & 0.626 ($\pm$0.136) \\
\cmidrule{2-6}
 & \multirow{2}{*}{Test} & MAPE & 8.568 ($\pm$14.37) & 8.518 ($\pm$15.85) & 8.888 ($\pm$15.30) \\ 
 & & Pearson R &0.351 ($\pm$0.238) &  0.354 ($\pm$0.239) &  0.295 ($\pm$0.257) \\
\cmidrule{2-6}
 & \multicolumn{2}{c}{Selected (\%)} & 34.9 & 40.4 & 24.7 \\ \midrule
 \multirow{6}{*}{Setting 2} & \multirow{2}{*}{Train} & MAPE & 0.229 ($\pm$0.257) & 0.249 ($\pm$0.251)
 & 0.257 ($\pm$0.252) \\
 & & Pearson R & 0.855 ($\pm$0.144) & 0.742 ($\pm$0.145) & 0.744 ($\pm$0.115) \\
\cmidrule{2-6}
 & \multirow{2}{*}{Test} & MAPE & 0.704 ($\pm$0.811) & 0.475 ($\pm$0.612)
& 0.477 ($\pm$0.538)  \\ 
 & & Pearson R & 0.420 ($\pm$0.189) & 0.512 ($\pm$0.181) & 0.466 ($\pm$0.183) \\  
\cmidrule{2-6}
 & \multicolumn{2}{c}{Selected (\%)} &  40.3 & 25.1 & 34.6 \\ \bottomrule
\end{tabular}
\label{results}
\end{table}

\begin{figure}
    \centering
    \includegraphics[width=0.95\textwidth]{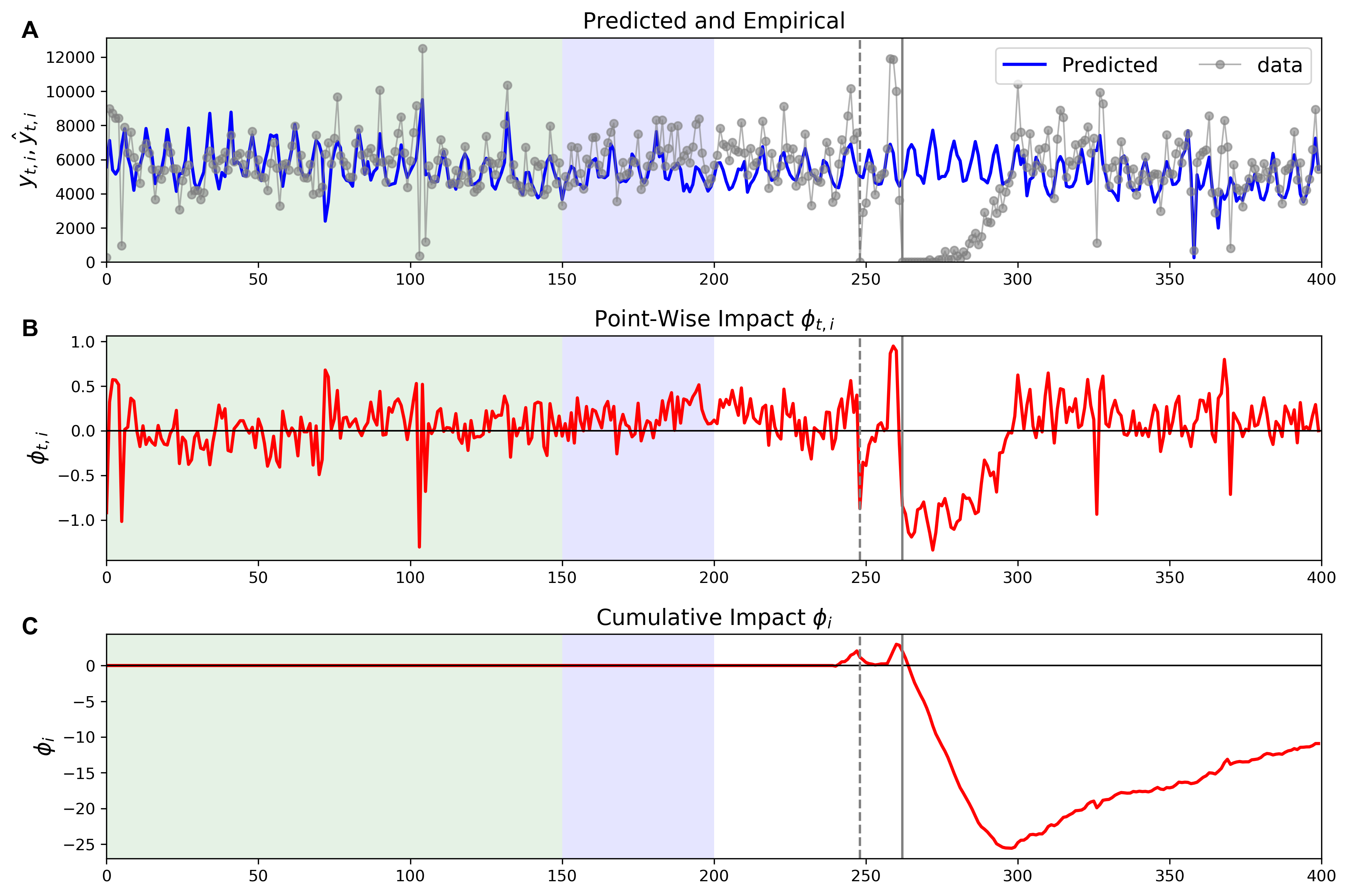}
    \caption{Example of how the disaster impact is quantified. (A) Predicted and actual observed daily visit patterns for a randomly selected business. (B) Point-wise impact $\phi_{t,i}$, and (C) cumulative impact $\phi_i$ of the disaster.}
    \label{sample}
\end{figure}

Figure \ref{sample} shows an example of how the the disaster impact is quantified. 
As shown in panel (A), we first predict the counterfactual daily visit counts after the disaster (blue plot) using the best performing model identified in the model validation experiment. 
Then, as shown in (B), we calculate the point-wise disaster impact $\phi_{t,i}$, by subtracting the observed daily visit count sequence from the predicted sequence and normalizing it by the pre-disaster mean daily visits. 
The cumulative disaster impact $\phi_i$ can be calculated by aggregating the point-wise disaster impacts over time. 
Panel (C) shows the cumulative disaster impact over time from the time of the landfall of the hurricane. 
In this particular business, we observe a significant negative impact until around day 300 with around $\phi_i = -25$, meaning that by day 300, this business lost a 25 business-as-usual days worth of customers due to the hurricane. 
We actually see positive impacts of the hurricane before the 2 hurricanes, however the positive impacts are significantly negated by the negative impacts. 
Gradually after 1 month from the hurricane landfall, we see an increase in visits compared to pre-disaster levels, which decrease the negative disaster impact.  
As a result of the BSTS modeling, we are able to obtain the quantified disaster impact for each of the businesses in Puerto Rico over time. 
In the next section, we analyze the obtained results to further understand which business categories in which locations suffered disaster impact in Puerto Rico after Hurricane Maria. 

\section*{Analysis of Estimated Business Resilience}
Now, using the BSTS method for predicting the counterfactual business performances, we quantitatively analyze the resilience of businesses after Hurricane Maria and answer the following questions: 
\begin{enumerate}
    \item How does the disaster impact evolve over time, and do the temporal patterns vary across business categories and locations? 
    \item Can we explain why we observe such heterogeneity in disaster impacts across businesses in Puerto Rico? 
\end{enumerate}
Since it was revealed that the optimal prediction models varied across different businesses in the Model Validation Section, we use the best performing model out of the three (either no covariate, average NY trend as covariate, or specific NY business trend as covariate) to predict the counterfactual visit time series for each of the businesses in Puerto Rico.

\subsection*{Quantifying Disaster Impact Patterns to Businesses}
\begin{figure}
    \includegraphics[width=0.9\textwidth]{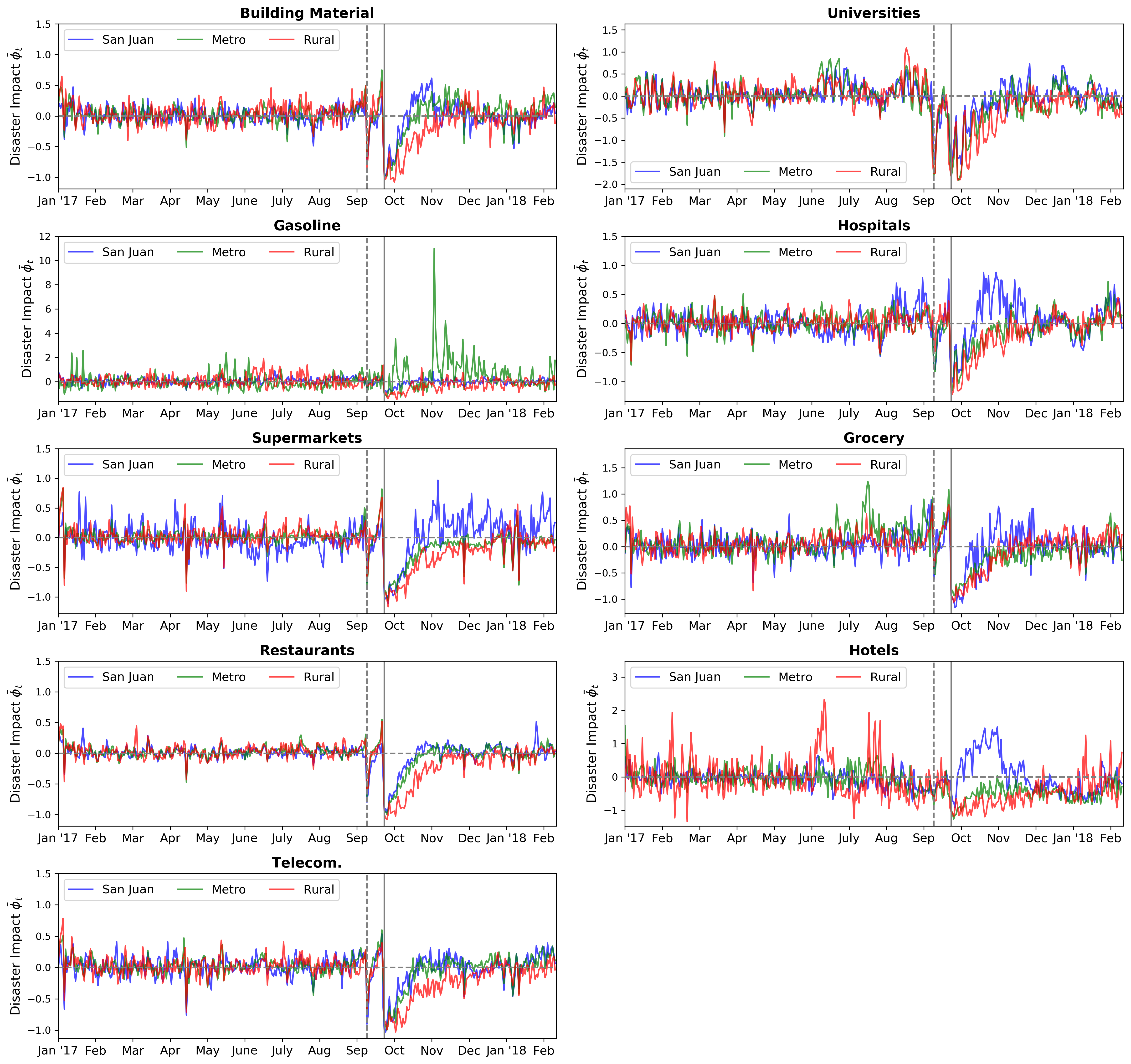}
    \caption{Point-wise disaster impacts across different business categories and business locations.}
    \label{timeseries}
\end{figure}

To answer the first research question, we aggregate the disaster impacts over the time horizon by business category and business location (San Juan Municipio, Metropolitan Area, Rural Area, shown in Figure \ref{map}B). 
Figure \ref{timeseries} shows the longitudinal point-wise disaster impact, which is the difference between the actual and the predicted business performances across time ($\phi_t = y_t - \hat{y_t}$) for all nine business categories. 
Negative values of $\phi_t$ would mean that the disaster had a negative impact on businesses, resulting in loss of customers, while a positive $\phi_t$ would mean that the number of customers increased due to the impact of the disaster. 
In each panel, the disaster impacts to businesses in the three regions are separately shown in blue (San Juan Municipio), green (Metropolitan Area), and red (rural area).
The vertical lines show the timings of the two hurricanes (dotted: Hurricane Irma, solid: Hurricane Maria).  

Several interesting observations can be made from these visualizations. 
First, we observe common trend across several business categories, where all three regions experience negative impact right after Hurricane Maria, and then the businesses in the urban areas recover quicker compared to those in rural areas.
This intuitive trend can be observed in various business categories including building materials, supermarkets, restaurants, telecommunications, and grocery stores. 
Second, we see a significant increase in gasoline stations in metropolitan areas (green) after Hurricane Maria. 
This reflects the high travel demand from the rural areas towards the metropolitan areas in the island due to evacuation mobility \cite{yabe2019mobile}. 
Third, in some business categories such as hospitals and hotels, we see an increase in visits after the hurricanes compared to before, especially in the San Juan region (blue). 
An increase in hospital visits reflect the large number of injuries and casualties caused by the flooding and severe winds caused by the hurricane. 
Significant increase in visits to hotels in San Juan reflect the large number of residents who evacuated from the rural areas in Puerto Rico to the capital city, which agrees with previous studies that observe the influx of population movements in San Juan from the suburban and rural areas of the island \cite{yabe2019mobile}. 
Minor details are captured in the figures as well, for example, how weekly fluctuations are estimated more vividly in universities (students do not attend classes on weekends) compared to other business types, and also how the impacts of Hurricane Irma, although minimal compared to Hurricane Maria, are captured in the time series data. 

\begin{figure}
    \centering
    \includegraphics[width=0.95\textwidth]{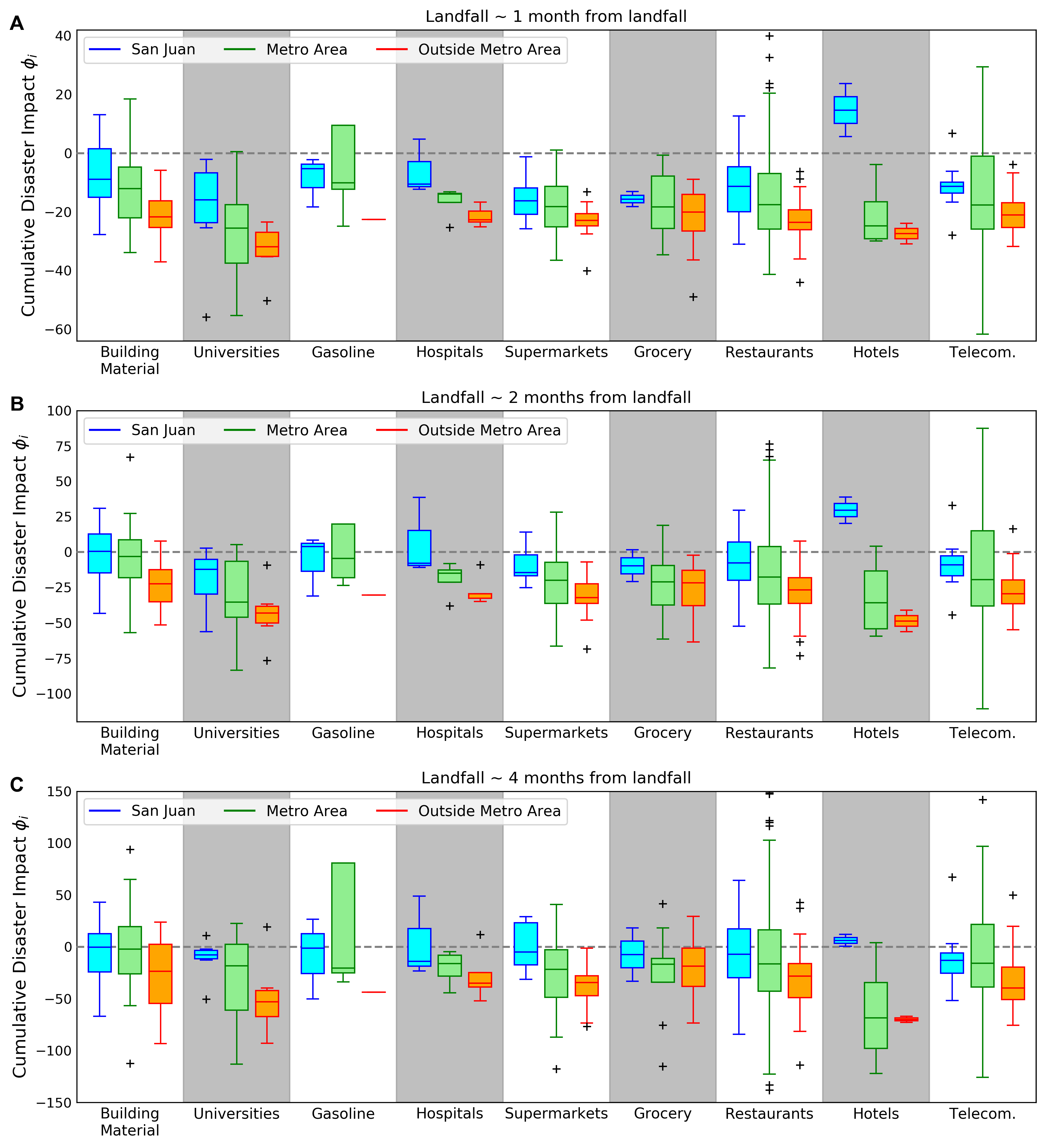}
    \caption{Cumulative disaster impacts across different business categories and regions. Results are shown for different aggregation time thresholds, including (A) landfall to 1 month from landfall, (B) until 2 months from landfall, and (C) until 4 months from landfall.}
    \label{boxplots}
\end{figure}

To further understand the impact of Hurricane Maria on the businesses, we computed the cumulative disaster impact ($\phi = \sum_t \phi_t$) for each business in each region. 
The cumulative disaster impacts are shown in Figure \ref{boxplots} for three different aggregation time thresholds, including (A) landfall to 1 month from landfall, (B) until 2 months from landfall, and (C) until 4 months from landfall.
For each business category, the cumulative disaster impacts are aggregated by regions, with the same color coding as Figure \ref{timeseries}.
The numbers of $\phi_i$ should be interpreted as ``the number of business-as-usual days worth of impact''.
For example, building material businesses in San Juan experienced a median disaster impact of $\phi = -10$ during the first month. 
This indicates that the building material businesses in San Juan lost 10 days worth of customers who were supposed to visit if the disaster did not occur. 
Most of the regions and business categories experience a negative impact in the first month, except for hotels in San Juan. 
We also clearly observe the urban-rural disparity in disaster impacts across many of the business categories across all three temporal thresholds. 
However, the urban-rural gap gradually closes down as time passes, and in many of the industries we observe little differences by 4 months from landfall (e.g. building material, grocery stores, restaurants, and telecommunications). 

Although the general patterns show consistent insights such as the urban-rural disparity, larger impact right after the landfall, and differences in disaster impacts across business categories, we are not able to delineate the effects of each characteristic on disaster impacts. 
In the next section, we attempt to reveal the impacts of business characteristics on the observed disaster impacts, by applying a hierarchical Bayesian modeling technique (ref).  

\subsection*{Delineating the Effects on Disaster Impact}
Although the results shown in the previous section revealed various patterns and correlations, the quantified disaster impacts were all conditioned on various features including the business characteristics (e.g. business category and location) and disaster characteristics. 
To delineate such effects and to understand the resilience of different business types, we apply a hierarchical Bayesian model approach. 
Hierarchical Bayesian models (HBMs) allow us to flexibly model the group-level effects on the estimand by introducing hyper prior distributions on the model parameters.
This is a significant difference from regular linear regression models which can only either i) assign one global parameter for all groups, or ii) estimate parameters separately for each group. 
For further details on the advantages of HBMs, readers should refer to \cite{gelman2013bayesian}. 

To estimate the cumulative disaster impact of all businesses, we construct the HBM as the following:
\begin{eqnarray}
    \left\{
    \begin{array}{c}
    \phi_i \sim N(\beta X_i + \delta_{r(i)} + \gamma_{c(i)}, \sigma^2) \\
    \delta_{r(i)} \sim N(0,\tau_{\delta}^2), \ \ \ \forall r \in \{0,1,2\} \ \ \ \# region \\
    \gamma_{c(i)} \sim N(0,\tau_{\gamma}^2), \ \ \ \forall c \in \{0,1,2,3,4,5,6,7,8\} \ \ \  \# category \\
    \beta, \sigma, \tau_\delta, \tau_\gamma \sim Cauchy(0, 2.5)
    \end{array}
    \right.
\end{eqnarray}
where, $r(i) \in \{0,1,2\}$ and $c(i) \in \{0,1,2,3,4,5,6,7,8\}$ denote the region index and category index for business $i$. 
We assume that the cumulative disaster impact on business $i$, denoted by $\phi_i$, can be modeled as a linear sum of the effects of exogenous features $X_i$ (which include pre-disaster business mean visits, housing damages caused by the disaster), regional effects $\delta_{r}$, and business categorical effects $\gamma_{c}$. 
The model is Bayesian in the sense that the model parameters ($\beta, \delta, \gamma$) all have priors, and the model is also hierarchical since the hyper-parameters in the prior distributions ($\tau_\delta, \tau_\gamma$) come from another higher level distribution.
We assume that the hyper-parameters are drawn from weakly informative priors (Cauchy distribution). 
The hierarchical prior distributions allow us to model the dependencies across different groups (regional groups and categorical groups). 

The model was implemented using \texttt{stan}, which performs sampling using Hamiltonian Monte Carlo method, and was coded on the \texttt{Pystan} package. 
Sampling was performed for 20,000 iterations with the first 1,000 used as warm-up. 
Thus, in total 19,000 samples were drawn for each parameter.
The sampling was ran on a regular laptop computer with an Intel i7 processor with 3.30GHz, and 8GB of RAM. 
The sampling took less than 5 minutes in total, which was much faster than the BSTS model due to the small number of parameters. 
The mixing of the sampling was effective, where $\hat{R}$ values were extremely close to 1.0000 ($\pm 0.0001$) for all parameters. Effective sample sizes were all significantly large, where the least was 5143. 

\begin{figure}
    \centering
    \includegraphics[width=0.9\textwidth]{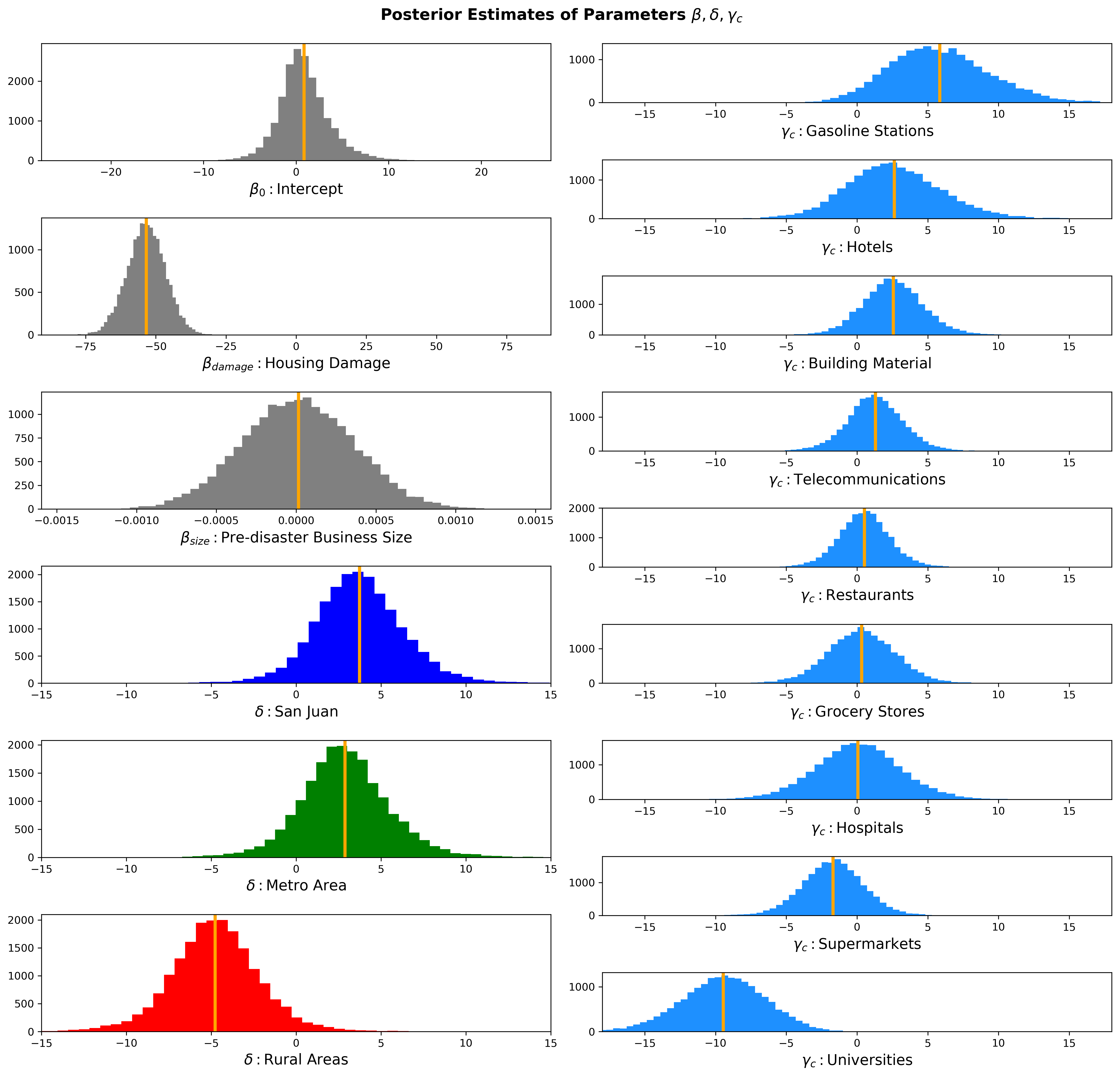}
    \caption{Posterior estimates of model parameters of the hierarchical Bayesian model. Using the model, we are able to understand the business impacts that each business category experiences conditioned on other factors such as housing damage rates, pre-disaster business sizes, and locations.}
    \label{posterior}
\end{figure}

Figure \ref{posterior} shows the posterior estimates of the model parameters in the hierarchical Bayesian model ($\beta, \delta, \gamma$).
The housing damage observed in the county of the business location had a significantly negative effect on the cumulative disaster impact, which was very intuitive. 
On the other hand, the intercept as well as the pre-disaster business size, which was measured by the mean visits to each business in the first 150 days of the observation (Hurricanes Irma and Maria struck on days 248 and 262), had no effect on the disaster impact. 
This contradicted previous studies which claim that business sizes have significant impact on the recovery of businesses \cite{sydnor2017analysis}. 
However, the study did not have detailed information on the category of the business (only whether or not the business was in the service sector). 
The effect of the business category may have negated the effect of pre-disaster business size. 
The estimated location effect agreed with our previous analyses (Figures \ref{timeseries} and \ref{boxplots}), showing that urban businesses had less negative disaster impact than rural ones. 
By delineating all of these effects, we are able to estimate the business impacts that each business category experiences, conditioned on other factors such as housing damage rates, pre-disaster business sizes, and locations.
The estimated effects ($\gamma$ parameter estimates) are shown in the right column of Figure \ref{posterior}. 
This shows that gasoline stations, hotels, building material, and telecommunications had positive disaster impacts, meaning that people visited these locations after the disaster more than before. 
This agrees with various news articles and studies that raise evidence of people rushing to purchase gas \cite{gas} and evacuating and staying in hotels \cite{hotel}.
This also reflects the household recovery process, where people purchase building materials for rebuilding homes and visits telecommunication companies to fix their mobile devices for internet connectivity. 
On the other hand, universities and supermarkets had a significant negative disaster effect. 
Again, this agrees with closures of universities in Puerto Rico after Hurricane Maria \cite{univ} and news articles pointing out under-supply in supermarkets after the disaster \cite{stores}. 

\section*{Discussion}
In this study, we used business visit data collected from mobile phone trajectories in Puerto Rico and New York to quantify the causal impact of Hurricane Maria on businesses in Puerto Rico.
Using the Bayesian Structural Time Series (BSTS) model, we predicted the counterfactual (what if the disaster did not happen) daily visit counts to businesses in Puerto Rico, and computed the point-wise disaster impact as well as the cumulative disaster impact of Hurricane Maria. 
Performances of the BSTS models were evaluated, and was found that whether the covariates (information of daily visit counts of businesses not affected by the disaster) positively contributes to the prediction accuracy varied across businesses. 
Furthermore, the estimated disaster impacts were analyzed using hierarchical Bayesian models to understand the effects of various business characteristics on disaster impacts. 

The findings in this study should be considered in the light of some limitations. 
First, the assumptions we make on the data to estimate business performance has several limitations. 
We used daily visit data as a proxy to estimate the performances of businesses in this study. 
However, we note that this approximation only holds when the business is a business-to-customer (B2C) type. 
If the main flow of transactions of the business is with other companies, mobile phone data would not be an appropriate data source for analysis. 
This is why we limited the study to 9 business categories that all usually have customers visit their stores to make profit. 
Moreover, although a previous study showed that the number of visits estimated from mobile phone data correlates well with actual business performances through the case study of a coffee chain \cite{gurun}, we could not fully validate this for other businesses in this study due to lack of corporate finance data. 
Further investigation on the relationships between business performances and customer visits after disasters would be worthy of investigation in future research. 
Second, although our study was able to produce more spatio-temporally granular and scalable analysis and estimations of disaster impacts on businesses compared to past studies using surveys, our study did miss some of the advantages of survey studies.  
Previous studies have revealed that more detailed characteristics of the businesses such as years of operation, number of employees, and age of the owner all affect the recovery performance after disasters \cite{sydnor2017analysis}.
Due to the limitations in data collection, we were not able to include such covariates in the hierarchical Bayesian model in the latter section of the analysis. 
Further efforts in combining different data sources (e.g. mobile phone data and survey data) to complement eachother would be a very interesting research direction. 
Third, from a methodological point of view, one could apply a more complex method to select and generate covariates for predicting future daily visit counts. 
In this study, we applied a heuristic approach in choosing the covariates for prediction, where the business that was most highly correlated with the target business daily counts was selected. 
Empirical validation showed that in some cases, using the covariate decreased the prediction accuracy. 
Efficient algorithms to detect and select appropriate covariates for future time series prediction would be of future research interest.
Finally, the analysis was performed for only Puerto Rico, thus the findings may not be generalizable to different disasters and locations, given the unique geographic and political characteristics of Puerto Rico. 
Expanding this analysis and comparing insights across different cities and disasters could generate more concrete insights and implications. 

We finally discuss how the methods, analysis, and findings presented in this study may be applied in disaster management and policy making.
As mentioned in the introduction, surveys have been the primary data source to estimate economic losses after disasters for policy makers. 
On the other hand, large scale mobility datasets have been more common in the decision making processes in various domains including epidemic control, traffic management and disaster relief. 
This study lays out an example of how such large scale mobility data can be used for i) post-disaster assessment and monitoring, ii) economic cost estimation, and iii) developing relief supply allocation strategies. 
As shown in Figures \ref{sample}, \ref{timeseries}, and \ref{boxplots}, we are able to quantitatively monitor the negative impact and recovery of each business using the models in this paper. 
It is not technically difficult to detect businesses that are struggling to recovery after the disaster, and carry out assistance programs for those businesses. 
Moreover, the estimated point-wise or cumulative disaster impacts can be multiplied with the average money spent per customer, to easily calculate the daily or total economic loss for each business. 
We can also identify the business categories that have not recovered in each region to develop strategies for allocating relief supplies, for example, for distributing gasoline across the island. 

\section*{Conclusion}
Quantifying the economic impact of disasters to businesses is crucial for disaster relief and preparation. 
The availability of large scale human mobility data enables us to observe daily visit counts to businesses in an unprecedented spatio-temporal granularity. 
In this work, we presented a methodology to estimate the causal impact of disasters to businesses from mobile phone location data, using a Bayesian modeling framework. 
The methodology was used to quantify the causal impact of Hurricane Maria on businesses in Puerto Rico. 
The estimation results provide insights on what types of businesses, located where, are able to recovery quickly after the hurricane. 
Such insights could assist policy makers during disaster preparation and relief processes.


\begin{backmatter}

\section*{Competing interests}
The authors declare that they have no competing interests.

\section*{Author's contributions}
TY, YZ, and SVU designed research, TY and SVU analyzed data, TY and YZ implemented models, and TY, YZ, and SVU wrote the paper.

\section*{Acknowledgements}
We thank Safegraph for preparing the mobile phone GPS data used in this study.
T.Y. is partly supported by the Doctoral Fellowship provided by the Purdue Systems Collaboratory. The work of T.Y. and S. V. U. is partly funded by NSF Grant No. 1638311 CRISP Type 2/Collaborative Research: Critical Transitions in the Resilience and Recovery of Interdependent Social and Physical Networks.  

\section*{Data Availability}
Mobile phone location data are proprietary data owned by private companies.
Although such data are not available for open access due to the users' privacy, we will obtain permission to post processed data that are sufficient to reproduce the results obtained in this study.
Data collected from other sources are available from official documents that are openly accessible.  
  

\bibliographystyle{bmc-mathphys} 
\bibliography{bmc_article}      

\end{backmatter}
\end{document}